\documentclass{egpubl}
\usepackage{sca2024}
\usepackage{amssymb}
\usepackage{amsmath}
\usepackage{multirow}
\usepackage{url}

\SpecialIssuePaper         

\CGFccby

\usepackage[T1]{fontenc}
\usepackage{dfadobe}  

\usepackage{cite}  
\BibtexOrBiblatex
\electronicVersion
\PrintedOrElectronic
\ifpdf \usepackage[pdftex]{graphicx} \pdfcompresslevel=9
\else \usepackage[dvips]{graphicx} \fi

\usepackage{egweblnk} 

\usepackage{xcolor}

\newcommand{\orange}[1]{\textcolor{black}{#1}}

\title[Diffusion-based Human Motion Style Transfer with Semantic Guidance]%
      {Diffusion-based Human Motion Style Transfer with Semantic Guidance}

\author[Hu et al.]
{\parbox{\textwidth}{\centering Lei Hu$^{1,2}$\orcid{0000-0001-8938-5071}, 
Zihao Zhang$^{1}$\orcid{0000-0001-6859-7518}, Yongjing Ye$^{1}$\orcid{0000-0002-1027-3382}, Yiwen Xu$^{1,2}$\orcid{0009-0002-6849-4046} and Shihong Xia\thanks{Corresponding author: xsh@ict.ac.cn}$^{1,2}$\orcid{0000-0002-7228-9646}
        }
        \\
{\parbox{\textwidth}{\centering $^1$Institute of Computing Technology, Chinese Academy of Sciences\\
            $^2$University of Chinese Academy of Sciences
       }
}
}

\begin{document}


\maketitle
\begin{abstract}
3D Human motion style transfer is a fundamental problem in computer graphic and animation processing. Existing AdaIN-based methods necessitate datasets with balanced style distribution and content/style labels to train the clustered latent space. However, we may encounter a single unseen style example in practical scenarios,  but not in sufficient quantity to constitute a style cluster for AdaIN-based methods. Therefore, in this paper, we propose a novel two-stage framework for few-shot style transfer learning based on the diffusion model. Specifically, in the first stage, we pre-train a diffusion-based text-to-motion model as a generative prior so that it can cope with various content motion inputs. In the second stage, based on the single style example, we fine-tune the pre-trained diffusion model in a few-shot manner to make it capable of style transfer. The key idea is regarding the reverse process of diffusion as a motion-style translation process since the motion styles can be viewed as special motion variations. During the fine-tuning for style transfer, a simple yet effective semantic-guided style transfer loss coordinated with style example reconstruction loss is introduced to supervise the style transfer in CLIP semantic space. The qualitative and quantitative evaluations demonstrate that our method can achieve state-of-the-art performance and has practical applications. The source code is available at \url{https://github.com/hlcdyy/diffusion-based-motion-style-transfer}.


\end{abstract}  
\section{Introduction}
\begin{figure}[ht]
    \centering
    \includegraphics[width=\linewidth]{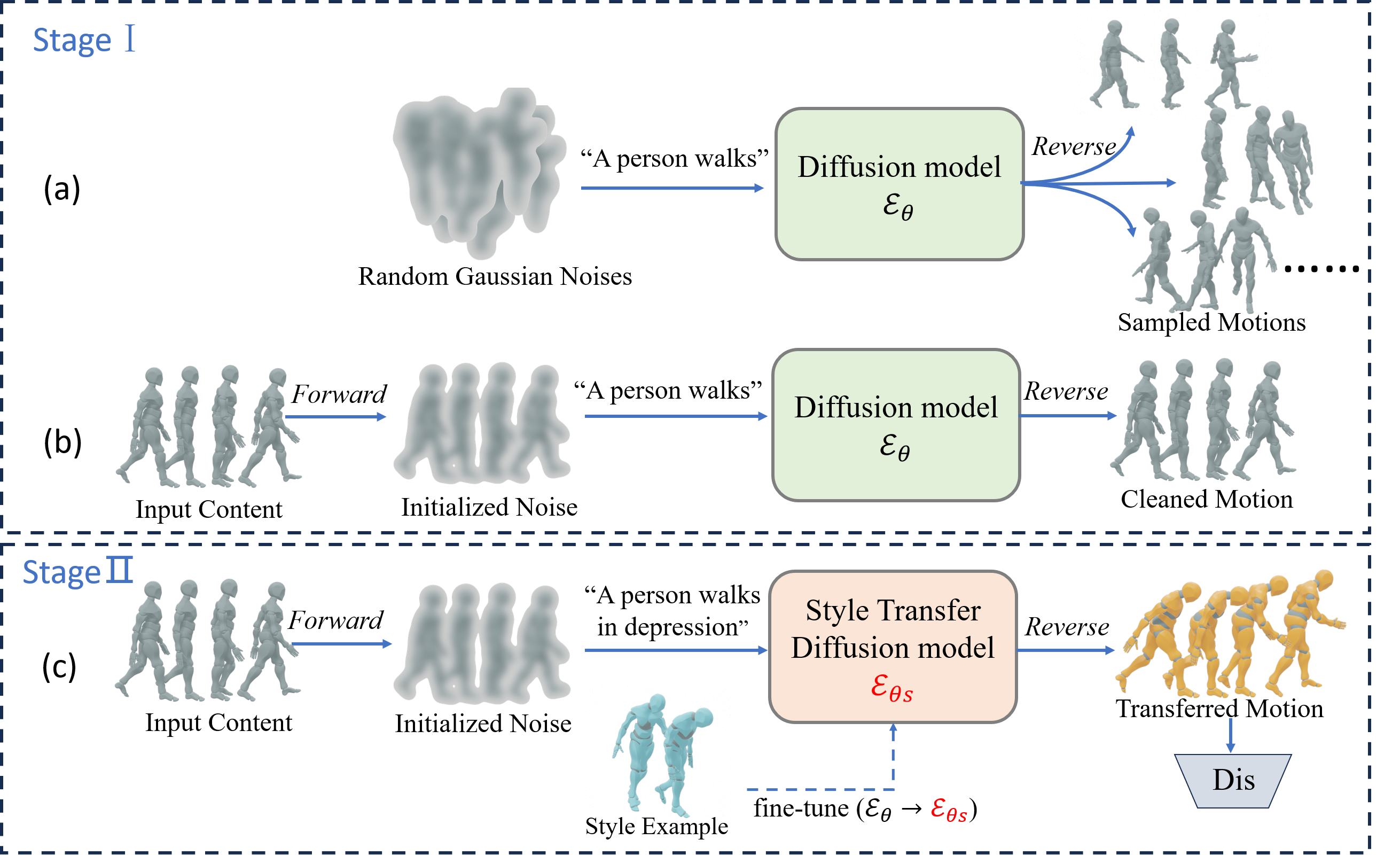}
    \caption{\orange{The concept figure of our framework.} (a) The pre-trained text-to-motion generative model; (b) The motion-to-motion translation model based on diffusion; (c) The fine-tuning for human motion style transfer. }
    \label{fig:concept}
\end{figure}

The stylized human motion can greatly enhance the artistic tension of the character, making it more lively and expressive in animation, gaming, and virtual reality. Currently, the primary way to acquire 3D human motion is motion capture. Nevertheless, it is infeasible to capture all of the infinite stylistic variations, and the capture relies on skilled actors, thereby restricting the scale of current style motion datasets. Consequently, human motion style transfer, which aims to transfer the style from a reference style example to other content motions, emerges as a compelling technique to enrich the human motion variations.

It is challenging to provide a precise mathematical definition of motion style since variations in both motion content and style are manifested in low-level motion features, such as joint rotations and global velocity. Therefore, style transfer, as a non-trivial problem, has been widely studied for decades. Prior approaches~\cite{hsu2005style,ma2010modeling,xia2015realtime} typically model style by examining differences between motion examples. However, it is not easy to obtain paired motion segments manually, and the laborious correspondence and preprocessing (e.g., aligning motion pairs through dynamic temporal warping~\cite{rabiner1993fundamentals}) hinders the model's ability to scale incrementally. Recently, AdaIN-based methods~\cite{aberman2020unpaired, jang2022motion} have emerged as the predominant approach for style transfer, offering the capability to disentangle content and style for unpaired motions and to cluster style features in the latent space. Nevertheless, AdaIN-based techniques necessitate datasets with balanced style distributions to train the clustered latent space. In practical scenarios, however, we may only have access to a limited number of stylistic examples (or even just a single clip) compared to a vast amount of neutral motion. Additionally, the AdaIN-based methods require re-training the model when confronted with unseen action types or styles that fall outside the dataset's distribution. Thus, the key issue to be addressed in human motion style transfer is: Given only one unseen style example, how to obtain a style transfer model capable of transferring a variety of content motions by few-shot learning.

Facing the challenge above, in this paper, we introduce a two-stage framework to split the human motion modeling and style transfer, as illustrated in Figure~\ref{fig:concept}. In stage \uppercase\expandafter{\romannumeral1}, \textbf{(a)} we pre-train a diffusion-based text-to-motion generative model $\varepsilon_\theta$ on a text-motion dataset. This model, functioning as a generative prior, can sample from Gaussian noise and generate diverse natural human motions from scratch under text guidance. We chose the text-motion generative model as a prior because the text prompts can provide more motion information relative to the content/style labels in the subsequent style transfer learning. \textbf{(b)} If we add noise to the input content motion by the forward process and denoise it by the reverse process of the diffusion model, the pre-trained text-to-motion generative model $\varepsilon_\theta$ can be seamlessly converted into a motion-to-motion translation model. And the noising and denoising processes inevitably introduce some variations to the input motion.

Given that motion styles can be viewed as special variations of human motions, we are inspired to regard the reverse process of the diffusion model as the style translation process. Specifically, in stage \uppercase\expandafter{\romannumeral2}, \textbf{(c)} given one style example, we fine-tune the diffusion network($\varepsilon_\theta \rightarrow \varepsilon_{\theta s}$) to enable the reverse process of $\varepsilon_{\theta s}$ to convert neutral noised motions into stylized motions. To guide the few-shot learning, we also employ a pre-trained motion-semantic discriminator to distinguish whether the transferred motion is semantically consistent with the content of the source motion and the target style. 

The proposed two-stage learning framework enables us to use a large text-motion dataset with unbalanced style distribution to pre-train a robust prior model. This ensures that the model is suitable for a wide range of content motion inputs during the second stage of style transfer fine-tuning. When presented with a style example, we only need to follow the stage \uppercase\expandafter{\romannumeral2} and fine-tune the $\varepsilon_\theta$ to obtain a corresponding style transfer model, without re-training on the whole dataset.  

During the fine-tuning, to perform the style transfer and preserve the original content of the input motion, we introduce two loss functions: the style example reconstruction loss $\mathcal{L}_{sr}$ and the semantic-guided transfer learning loss $\mathcal{L}_{s}$. Specifically, for $\mathcal{L}_{sr}$, we first use the generative prior $\varepsilon_{\theta}$ and motion in-painting technique to generate an aligned neutral motion corresponding to the style example. Subsequently, we enforce the style transfer diffusion model $\varepsilon_{\theta s}$ to translate the corresponding neutral motion into a motion resembling the style example. For $\mathcal{L}_{s}$, we use a pre-trained motion-semantic discriminator to assess the alignment of the output motions with the semantics of the original content and the style of the example motion in the CLIP~\cite{radford2021learning} space.

The contributions of our work can be summarized as follows:
\begin{itemize}
    \item We propose a novel style transfer framework, which regards the reverse process of the diffusion model as a style translation process.
    \item We propose a novel semantic-guided style transfer learning strategy, which can supervise human motion style transfer in the CLIP semantic space.
    \item Extensive experiments demonstrate that the proposed framework achieves unseen style transfer on only one style example through fine-tuning.   
\end{itemize}

\section{Related Works}
\orange{In this section, we will first review work related to human motion style transfer (\ref{ssec:style_transfer}) and then review techniques related to motion generation (\ref{ssec:motion_generation}).}
\subsection{Human Motion Style Transfer}\label{ssec:style_transfer}

\textbf{Example-based style transfer.}
Human motion style transfer has been a long-standing problem in the field of computer graphics for decades. The early example-based methods mostly regard the style transfer as a translation problem and employ the examples to learn the translation model~\cite{brand2000style,hsu2005style,wang2007multifactor, ikemoto2009generalizing,ma2010modeling, xia2015realtime,yumer2016spectral}. 
~\cite{hsu2005style} learn the mapping between motions with identical content but different styles by a linear time-invariant system, which can then be used to transfer a new motion to the learned style in a real-time manner. 
~\cite{ma2010modeling} employed a Bayesian network to effectively model the relationship between user-specified parameters and latent variation parameters, which capture the style variation. 
~\cite{xia2015realtime} employs a series of local mixtures of auto-regressive models to learn style differences and achieve notable results on unlabeled, heterogeneous motion data.~\cite{yumer2016spectral} proposes a spectral style transfer to circumvent the costly and time-consuming requirement of style databases. Typically, the matching motion pairs necessary for these approaches are either manually provided or searched in motion databases.

\noindent\textbf{Human motion style transfer in latent space.}
In recent years, with the development of deep learning, various advanced methods have been employed on human motion style
transfer tasks.~\cite{holden2016deep, holden2017fast}, ~\cite{du2019stylistic} and ~\cite{wang2021cyclic} construct the motion manifold by the convolutional network, CVAE, and CycleGAN~\cite{zhu2017unpaired}, respectively. They explicitly define the Gram matrix~\cite{gatys2016image} of the deep features as the human motion style, while the feature as content. The concept proposed by~\cite{du2019stylistic} bears a resemblance to our approach, which uses a few style examples and a neutral motion dataset to achieve style transfer. Nonetheless, their transfer is limited to locomotion. 

Some methods integrate style modeling with motion generation. ~\cite{mason2018few, mason2022real,wen2021autoregressive} regard the motion style as a condition in real-time motion synthesis. ~\cite{mason2018few} introduce residual adapters to adjust the weights of PFNN~\cite{holden2017phase} to transfer the locomotion style during the real-time controlling. ~\cite{wen2021autoregressive} proposes a generative flow model and uses the style feature as a condition to synthesize the stylized human motion in an autoregressive manner. ~\cite{mason2022real} extend the work~\cite{mason2018few} and proposes a style modulation network to extract style information from motion clips and use it to modulate the hidden layers of the motion synthesis network. However, most of these methods are also restricted to style transfer for locomotion.

\noindent\textbf{AdaIN-based human motion style transfer.}
Due to the remarkable performance in image style transfer~\cite{huang2017arbitrary}, AdaIN has been extensively employed in the human motion style modeling~\cite{aberman2020unpaired, park2021diverse, jang2022motion, song2023finestyle}. It implicitly decouples the motion into content and style codes and enables style transfer through code recombination. This type of method can be regarded as a special case of human motion style transfer methods in latent space. ~\cite{aberman2020unpaired} is the first work to use AdaIN in the human motion style transfer task. ~\cite{jang2022motion} achieves style control over different body parts using part-based AdaIN.
FineStyle~\cite{song2023finestyle} proposes a fine-grained style transfer framework by introducing the content label and fusing it with style code in the latent space for more accurate modeling. ~\cite{ao2023gesturediffuclip} uses AdaIN to modulate the features in the latent diffusion module to achieve stylized gesture generation with different style prompts. The AdaIN-based approach has shown promising results in addressing style transfer between non-locomotion. However, the methods require a dataset with balanced style motion to learn the clustered latent space. Therefore, it is difficult to learn style information when given only one unseen style example.

Our framework does not require the dataset to be stylistically balanced when pre-training the generative prior, and in the fine-tuning stage, we need only one style example to generate a style transfer diffusion model that can be applied to a variety of content motion inputs.

\subsection{Human Motion Generation}\label{ssec:motion_generation}

Generating natural-looking human motion with user-specific constraints has always been a research focus in the field of human motion synthesis. As for traditional constraints, auto-regressive networks~\cite{holden2017phase, starke2020local, wang2019combining, ling2020character} have demonstrated flexible motion control results given joystick signals. As for sparse constraints such as VR trackers, physics-based motion tracking and generation methods~\cite{ye2022neural3points, winkler2022questsim} have achieved remarkable performance. Recently, there has been a substantial surge of interest in cross-modal constraints,  especially those with semantic prompts~\cite{petrovich2021action, zhong2022learning, athanasiou2022teach,Guo_2022_CVPR, jiang2023motiongpt, zhang2023generating, zhong2023attt2m}
The primary approach of these works focuses on utilizing appropriate generative models in latent space. For example, ACTOR~\cite{petrovich2021action} and TEACH~\cite{athanasiou2022teach} leverage transformer-VAE architecture to construct a mapping between label/text prompts and motions. Similarly, Vector-Quantized VAE(VQ-VAE) ~\cite{van2017neural}, which quantize continuous latent into discrete representations, 
demonstrate its ability on text-to-motion tasks~\cite{jiang2023motiongpt,zhang2023generating, zhong2023attt2m} combined with Generative Pre-trained Transformer~\cite{brown2020language}.

Unlike encoder-decoder frameworks, diffusion-based methods directly sample from the original motion space, providing a more stable training approach for motion generation. Denoising Diffusion Probabilistic Models ~\cite{ho2020denoising} combined with CLIP textual features is wildly used for text-driven motion synthesis tasks~\cite{tevet2023human, zhang2023remodiffuse, zhang2024motiondiffuse}. Among them, the conditioned-synthesis pipeline introduced by MDM~\cite{tevet2023human} showcases its flexibility in adapting to various new tasks, including physics constraints~\cite{yuan2023physdiff} and music-dance synthesis~\cite{tseng2023edge}. 
Moreover, PriorMDM ~\cite{shafir2024human} uses MDM as a generative prior and extends its ability on motion inpainting. SinMDM~\cite{raab2024single} employs the U-Net architecture in the diffusion model and uses only single-segment motion as training data. Particularly, they implement style transfer as a special case of harmonization. ~\cite{yin2024scalable} connects the noise spaces of different style motion based on Dual Diffusion Implicit Bridges~\cite{su2022dual} and ensures the preservation of motion content by keyframe manifold constraint gradients. In our work, inspired by the StyleDiffusion in image~\cite{wang2023stylediffusion}, we consider the reverse process of the diffusion model as a style translation process, which models the style difference between style-neutral examples. 

\section{Method}
In this section, we will first introduce the motion representation in Sec~\ref{ssec:representation}. Then, we review the preliminary knowledge of the diffusion models (Section~\ref{ssec:preliminary}). In Sec~\ref{ssec:G_and_D}, we will describe stage \uppercase\expandafter{\romannumeral1}: pre-training of generative model and motion-semantic discriminators. In Sec~\ref{ssec:fine-tune}, we will introduce stage \uppercase\expandafter{\romannumeral2}: the fine-tuning for style transfer. Finally, in Sec~\ref{ssec:inference}, we describe the style transfer at inference time and the post-process. 
\begin{figure*}[ht]
    \centering
    \includegraphics[width=\linewidth]{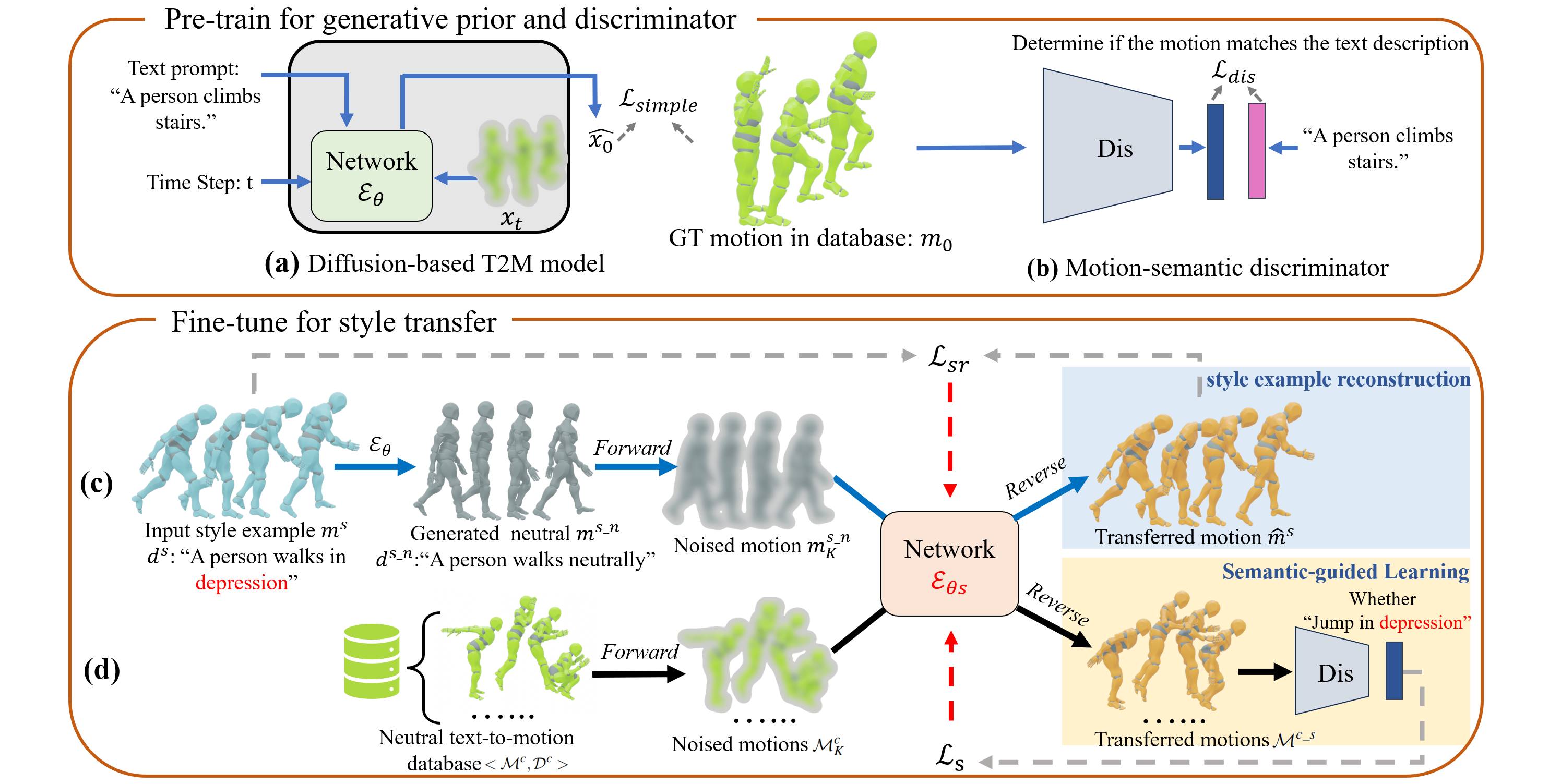}
    \caption{Overview of our learning process. In the pre-training stage, we train the diffusion-based T2M model $\varepsilon_{\theta}$ (a) as generative prior and a motion-semantic discriminator (b) for measuring the similarity between motion and textual prompts. In the fine-tuning stage, we employ the style example reconstruction loss (c) to utilize the style information provided by example $m^s$. Concurrently, we use the pre-trained motion-semantic discriminator to guide the transferred motion to match the semantics of the original content and the style of the example motion (d). Notably, input style example, generated neutral motion, content motion, and output transferred motion are colored light blue, grey, green, and yellow, respectively. }
    \label{fig:overview}
\end{figure*}
\subsection{Motion Representation}\label{ssec:representation}
A motion sequence $m\in \mathcal{M}$ is represented by a temporal set of $f$ poses. Each pose feature comprises the following three components, namely, the root joint vector $\mathbf{o}\in \mathbb{R}^4$, local joint positions relative to the root $\mathbf{p}\in \mathbb{R}^{(J-1)\times 3}$ and local joint rotations $\mathbf{r}\in \mathbb{R}^{J\times 6}$ relative to their parent joint coordinate frame, where $J$ denotes the number of joints.
Additionally, the root joint vector $\mathbf{o}$ includes the y-axis angular velocity $\mathbf{o}^r$, xz-plane linear velocity $\mathbf{o}^x, \mathbf{o}^z $, and absolute root height on the y-axis $\mathbf{o}^h$. 
The rotation representation in $\mathbf{r}$ is the 6D vector defined in the work~\cite{zhou2019continuity}\orange{, which is actually the first two column vectors of the rotation matrix}. The y-axis angular velocity $\mathbf{o}^r$ of the root joint is obtained from the change in the human-facing direction. The facing direction is calculated similarly to PFNN~\cite{holden2017phase}, which averages the vector between the hip joints and the vector between the shoulder joints and takes the cross product with the upward direction. 

To obtain facing-independent motion features, we further localize the $\mathbf{o}^x$, $\mathbf{o}^y$, $\mathbf{p}$, and the root joint rotation $\mathbf{r_0}$ based on the facing direction of the current pose. These processed features are concatenated to formulate $\mathbf{v} = [\mathbf{o},\mathbf{p},\mathbf{r}]$ as pose representation, and the motion $m$ of length $f$ can be expressed as $m=[\mathbf{v}_1, \mathbf{v}_2, ..., \mathbf{v}_f]$. We denote the style motion and content motion as $m^s$ and $m^c$, respectively. (The motion lengths of $m^s$ and $m^c$ can be different.)

\subsection{Preliminary of the Diffusion Models}\label{ssec:preliminary}

The diffusion models~\cite{ho2020denoising, song2021denoising} consist of two processes: Forward diffusion process and Reverse denoising process. The forward process is modeled as a Markov chain, where the Gaussian noise is gradually added to real data $m_0\in \mathcal{M}$ until $m_T$ is equivalent to an isotropic Gaussian noise.  For a single forward step $t$, the noising process can be formulated as follows (note the time step $t$ is not the same concept as the motion frame $f$):
\begin{equation}
    \orange{q(m_t|m_{t-1}) = \mathcal{N}(\sqrt{\alpha_t} m_{t-1}, (1-\alpha_t) I)}
\end{equation}
where \orange{$q$ is the conditional probability distribution. The} $\alpha_t\in(0, 1)$ are constant hyper-parameters and $\alpha_1 >\alpha_2>...>\alpha_T$, $I$ is an identity matrix. When $\alpha_t$ is small enough, $m_T$ can be approximated as a standard normal distribution $\mathcal{N}\sim(0, 1)$. With the reparameterization trick, we can sample $m_t$ at any $t$ step directly based on the original data $m_0$:
\begin{equation}
    m_t = \sqrt{\bar{\alpha}_t} m_0 + \sqrt{1-\bar{\alpha}_t} \epsilon \quad\quad \epsilon\sim\mathcal{N}(0, 1)
    \label{eq:noise_step}
\end{equation}
where $\bar{\alpha}_t$ is the cumulative product from $\alpha_1$ to $\alpha_t$. From the Equation~\ref{eq:noise_step}, it can be seen that the noised feature $m_t$ consists of an $m_0$ term and a random Gaussian noise $\epsilon$ term, and the effect of $m_0$ on $m_t$ gradually diminishes as $t$ increases. 

The reverse process is the inverse of the above process. The true transition probability $q(m_{t-1}|m_t)$ is approximated using a neural network $\varepsilon_\theta$. Instead of predicting the noise $\epsilon_t$ as originally proposed by~\cite{ho2020denoising}, we follow~\cite{ramesh2022hierarchical,tevet2023human} and directly predict the initial motion:
\begin{equation}
    \hat{m}_{0(t)} = \varepsilon_\theta(m_t, t, c) 
\end{equation}
Where the $c$ is the conditioning signal in conditioned motion synthesis. $\hat{m}_{0(t)}$ is the initial motion predicted from $m_t$ at $t$ step. 

\subsection{Pre-train for Generative Prior and Discriminator}\label{ssec:G_and_D}
\begin{figure}
    \centering
    \includegraphics[width=\linewidth]{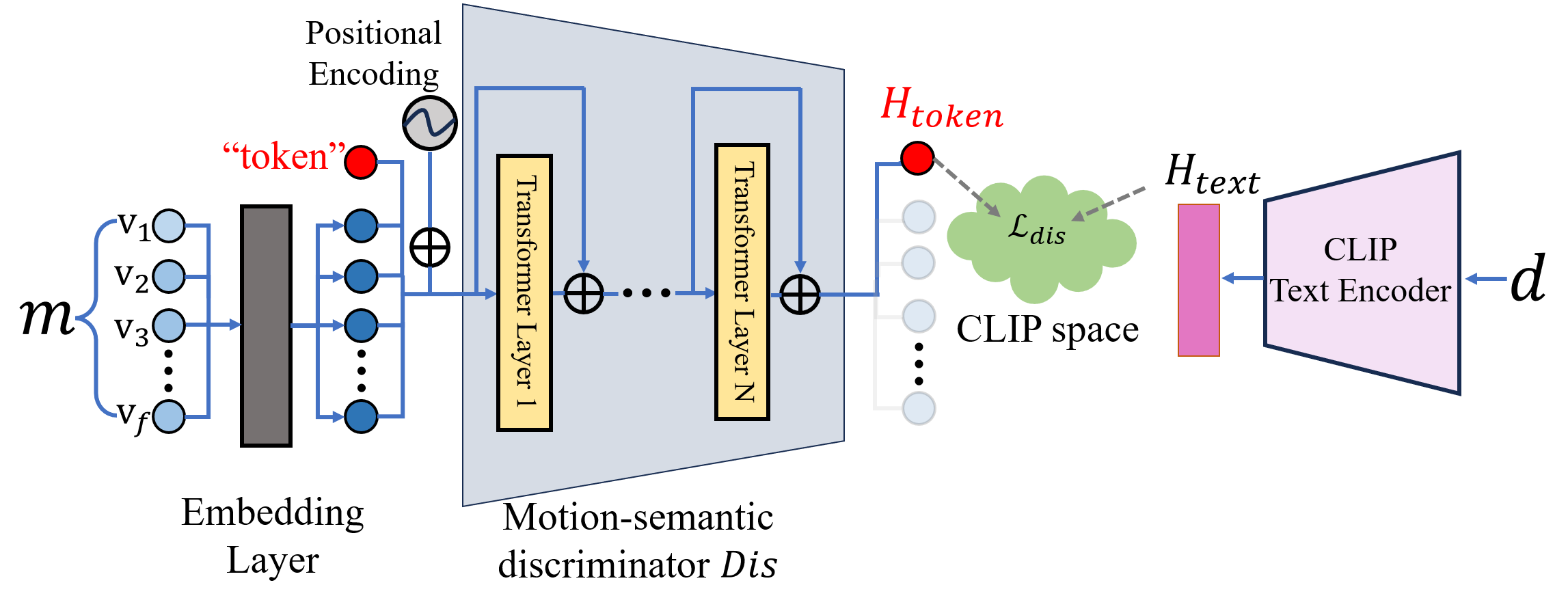}
    \caption{The architecture of our motion-semantic discriminator. We introduce a trainable "token" (short for "semantic token") for motion pooling and align the motion-semantic feature $H_{token}$ with the text feature $H_{text}$ in the CLIP space.}
    \label{fig:discriminator}
\end{figure}
As illustrated in Figure~\ref{fig:overview}, we will first pre-train a diffusion-based text-to-motion (T2M) model $\varepsilon_\theta$ as generative prior and a motion-semantic discriminator $Dis$ in stage \uppercase\expandafter{\romannumeral1}. 

Previously, autoencoders~\cite{holden2016deep, du2019stylistic} were commonly used to build motion manifolds for downstream tasks. While reconstruction of input in autoencoders can learn a good motion prior from the dataset, the statistical model only contains modal information about the 3D motion and no other control signals. The T2M generative model $\varepsilon_\theta$ allows us to use semantics as a condition to generate motion by sampling in the noise space, which is beneficial for our subsequent fine-tuning of style transfer.

\textbf{Pretraining of T2M generative model $\varepsilon_\theta$.} Given the motion-text pair $<m,d>$ in dataset, the training process of $\varepsilon_\theta$ is similar to~\cite{tevet2023human}. Text features produced by the CLIP text encoder are first added with time-step embeddings, and then concatenated with noised motion feature $m_t$ sampled by the forward process(see Equation~\ref{eq:noise_step}). Finally, the concatenated features are fed into the network $\varepsilon_\theta$ to clean the motion:
\begin{equation}
    \hat{m}_{0(t)} = \varepsilon_\theta(m_t, t, d)
\end{equation}

The $\varepsilon_\theta$ is trained by encouraging the prediction $\hat{m}_{0(t)}$ at each time step $t$ to be close to the initial signal $m_0$, i.e., $m$, the loss function can be written as:
\begin{equation}
    \mathcal{L}_{simple} = \mathbb{E}_{m, t\sim[1, T]}[\|m_0-\varepsilon_\theta(m_t, t, d)\|_2^2]
\end{equation}

\textbf{Pretraining of motion-semantic discriminator $Dis$.} Different from the original discriminator network~\cite{creswell2018generative} which only judges fake or real, our proposed motion-semantic discriminator aims to determine whether the motion matches the semantics of a given text. This judgment is achieved by comparing the cosine similarity between the motion features and the text features (see Figure~\ref{fig:overview} (b)).

The architecture of the motion-semantic discriminator is shown in Figure~\ref{fig:discriminator}. In order to process motion of different lengths, we follow~\cite{petrovich2021action, hu2023pose} and introduce a trainable parameter called "semantic token". The "semantic token" has the same dimension as the pose embeddings, and it will be concatenated with them and fed into the discriminator $Dis$ after position encoding. Among the output features, we only keep the feature $H_{token}$ at the position corresponding to the "semantic token" for pooling purposes. During pre-training, the pooled motion feature is encouraged to be aligned with its paired text feature in the CLIP space using contrast learning:
\begin{equation}
     \orange{\mathcal{L}_{dis} = \mathbb{E}_{m, d}[1 - \frac{Dis(m)\cdot E_{text}(d)}{max(\|Dis(m)\| \times \|E_{text}(d)\|, \delta)}]
    \label{eq:loss_dis}}
\end{equation}

Where $<m, d>$ is from the motion-text dataset and $E_{text}$ is the off-the-shelf CLIP text encoder. \orange{The loss $\mathcal{L}_{dis}$ measures the cosine similarity between vectors $Dis(m)$ and $E_{text}(d)$, where} $\delta$ is a small value to avoid division by zero, which is set to $10^{-6}$ in our setting.

\subsection{Fine-tune for Style Transfer}\label{ssec:fine-tune}
The overview of fine-tuning for style transfer is shown in the bottom half of Figure~\ref{fig:overview}. Given one style example $m^s$ and a prompt $d^s$ for its content and style, we aim to obtain a style transfer diffusion model $\varepsilon_{\theta s}$ related to $m^s$  through fine-tuning ($\theta\rightarrow \theta s$, different style examples will correspond to different fine-tuned model). After fine-tuning,  $\varepsilon_{\theta s}$ can transfer the style of $m^s$ to diverse neutral motions while preserving the original motion content. To achieve that, we introduce a style example reconstruction loss $\mathcal{L}_{sr}$ (see Figure~\ref{fig:overview}(c)) coordinated with a CLIP-based semantic-guided learning loss $\mathcal{L}_s$ (see Figure~\ref{fig:overview}(d)) to jointly fine-tune the network.

\textbf{\orange{Style} example reconstruction.} As shown in Figure~\ref{fig:overview}(c), in order to fully utilize the style information provided by the input example $m^s$, we design the style example reconstruction loss $\mathcal{L}_{sr}$. Specifically, we first obtain the neutral motion $m^{s\_n}$ corresponding to $m^s$, then we add noise to the neutral motion through the forward process and finally reconstruct the style example by the reverse process to learn the difference between $m^{s\_n}$ and $m^s$. 

To generate style-neutral motion pair $<m^s, m^{s\_n}>$, previous translation-based style transfer methods obtain matching motions by manually providing~\cite{hsu2005style} or by KNN search over a mocap dataset~\cite{xia2015realtime} and use DTW~\cite{rabiner1993fundamentals} or GTW~\cite{zhou2012generalized} to apply spatial and temporal warps to align the two motions. However, these procedures are tedious and time-consuming. In this work, we propose to use the pre-trained $\varepsilon_\theta$ as a statistical model to get the well-aligned motion pair $<m^s, m^{s\_n}>$. Specifically, we first add G-step noise to $m^s$ to obtain the noised motion $m^s_G$ to remove motion details by the forward process (see Equation~\ref{eq:noise_step}). During the reverse process, we use $m^s_G$ as the initial noised motion feature coordinated with a neutral text prompt $d^{s\_n}$ to generate paired natural motion $m^{s\_n}_{0(t)}$ at step $t$:
\begin{equation}
    m_{0(t)}^{s\_n} = \varepsilon_\theta(m_t^s, t, d^{s\_n}) \quad \text{where}~~ t: from ~G~ to~ 0
    \label{eq:mdm_denoising} 
\end{equation}

Where $d^{s\_n}$ is obtained by modifying $d^s$, i.e., we replace style-related vocabulary in $d^s$ while preserving the content-relevant descriptions. For instance, in Figure~\ref{fig:overview}(c), we substitute "neutrally" for "in depression" in $d^s$. This substitution can be achieved by AMR~\cite{liu2018toward} or a large language model like ChatGPT 3.5~\cite{chatgpt}. Upon obtaining $m_{0(t)}^{s\_n}$, we can infer the noised motion at step $t-1$ by the forward process, thus progressively generating final neutral motion $m^{s\_n}$, i.e., $m_{0(0)}^{s\_n}$.

In addition to prompt revisions, we incorporate the root trajectory of the style example projected onto the horizontal plane as the inpainting features for spatio-temporal alignment. Specifically, we extract the linear velocities of the root joint in $m^s$ frame-by-frame, denoted as $\orange{O_{hori}} = [\mathbf{o}_1^x, \mathbf{o}_1^z, \mathbf{o}_2^x, \mathbf{o}_2^z,...,\mathbf{o}_f^x, \mathbf{o}_f^z]$. After obtaining the denoised motion $m_{0(t)}^{s\_n}$ for each time step $t$ according to the Equation~\ref{eq:mdm_denoising}, we substitute the corresponding portion of linear velocity with the velocities from the style example $m^{s}$:
\begin{equation}
     m_{0(t)}^{s\_n}\orange{(O_{hori})} = m^{s}\orange{(O_{hori})}
\end{equation}

This inpainting acts similarly to DTW~\cite{rabiner1993fundamentals}, allowing the generated motion to be aligned with the style example $m^s$. Eventually, we take the final denoised motion $m_{0(0)}^{s\_n}$ as the generated neutral motion $m^{s\_n}$. In summary, the initial noise $m_G^s$ sampled by $m^s$ and the root trajectory inpainting ensure that the generated motion $m^{s\_n}$ is consistent with the content of $m_s$, while the neutral text prompt ensures that the generated motion is stylistically neutral.

After obtaining the style-neutral motion pair $<m^s, m^{s\_n}>$, we can encourage the reverse process of diffusion model $\varepsilon_{\theta s}$ to complete the task of style transfer by simply reconstructing the style example. Specifically, we first add K-step noise to $m^{s\_n}$ through the forward process, resulting in the noised motion $m_K^{s\_n}$. Then, we perform the reverse process to clean the motion:
\begin{equation}
     \hat{m}_{0(t)}^{s} = \varepsilon_{\theta s}(m_{t}^{s\_n}, t, d^s) \quad \text{where}~~ t: from ~K~ to~ 0
    \label{eq:style_denoising}
\end{equation}

Rather than predicting the motion $\hat{m}_{0(t)}^{s}$ to be identity to the $m^{s\_n}$, we encourage it to be close to $m^s$ for style transfer:
\begin{equation}
    \mathcal{L}_{sr} = \|m^s - \hat{m}_{0(t)}^{s} \|_2^2 \quad t: from ~K~ to~ 0
\end{equation}

\textbf{Semantic-guided learning.} As illustrated in Figure~\ref{fig:overview}(d), we resort to diverse neutral motions in the text-to-motion database $<\mathcal{M}^c, \mathcal{D}^c>$ and use semantic-guided style transfer loss $\mathcal{L}_s$ to simultaneously fine-tune the network $\varepsilon_{\theta s}$ along with $\mathcal{L}_{sr}$. Specifically, we add K-step noise to content motions $\mathcal{M}^c$ through the forward process, resulting in $\mathcal{M}^c_K$. \orange{Meanwhile}, we modify the set of neutral text prompt  $\mathcal{D}^c$ to $\mathcal{D}^{c\_s}$ with style vocabulary of $m^s$. \orange{Obtaining the set of stylized text prompts $\mathcal{D}^{c\_s}$ is similar to getting $d^{s\_n}$ in Equation~\ref{eq:mdm_denoising}.} For example, in Figure~\ref{fig:overview}(d), "A person jumps neutrally." is changed to "A person jumps in depression.".

We then feed the noised motions $\mathcal{M}^c_K$ into the network $\varepsilon_{\theta s}$ for denoising and style transfer. \orange{To make} the output transferred motions $\mathcal{M}^{c\_s}$ maintain the content of the original motions while having the style corresponding to $m^s$, we adopt the pre-trained motion-semantic discriminator $Dis$ in~\ref{ssec:G_and_D} to guide the fine-tuning. Specifically, the transferred motion \orange{$m^{c\_s}\in\mathcal{M}^{c\_s}$} and the \orange{stylized} text prompt \orange{$d^{c\_s}\in\mathcal{D}^{c\_s}$} are fed into $Dis$ and the CLIP text encoder $E_{text}$, respectively. The semantic-guided loss $\mathcal{L}_s$ is computed by the cosine similarity of the motion and text features:
\begin{equation}
    \orange{\mathcal{L}_{s} = \mathbb{E}_{m^{c\_s}, d^{c\_s}}[1 - \frac{Dis(m^{c\_s})\cdot E_{text}(d^{c\_s})}{max(\|Dis(m^{c\_s})\|\times \|E_{text}(d^{c\_s})\|, \delta)}]}
    \label{eq:semantic_loss}
\end{equation}

Finally, the style transfer fine-tuning loss is defined as a compound of $\mathcal{L}_{sr}$ and $\mathcal{L}_s$:
\begin{equation}
    \mathcal{L}_{total} = \lambda_{sr} \mathcal{L}_{sr} + \lambda_{s} \mathcal{L}_{s}%
    \label{eq:finetune_loss}
\end{equation}

\subsection{Style Transfer at Inference}\label{ssec:inference}
Given the fine-tuned style transfer diffusion model $\varepsilon_{\theta s}$, we can apply the style of $m^s$ to diverse content motions $m^c$. If the source motion $m^c$ is not stylistically neutral, we can first use the generative prior $\varepsilon_{\theta}$ to generate the corresponding neutral motion(similar to the style-neutral generation in Sec~\ref{ssec:fine-tune}) as an intermediate state. Then, the style transfer can be easily achieved by forward process ($m^c\rightarrow m^c_K$) and reverse process ($m^c_K\rightarrow m^{c\_s}$). 

\textbf{Global Velocity.} Global velocity is also a property of motion style, for example, the velocity should decrease when transferring the neutral walking to old walking. However, the translation of global velocity is a challenging task, especially when the root position of the style example is almost static (e.g. given old kicking as the style example). In fact, during the finetuning for the style transfer in Sec~\ref{ssec:fine-tune}, we learn the translation between temporal-aligned motions.  
Therefore, to reflect the style differences in the global velocity, we follow the~\cite{aberman2020unpaired} and perform a time warping on the global velocity based on the following velocity vector:
\begin{equation}
    \mathcal{U} = \frac{1}{f}\sum_{\tau}^f \max_{j\in J} u^j(\tau)
\end{equation}

Where the $u^j({\tau})$ is the magnitude of the local velocity of the $j$-th joint in time $\tau$. The $\mathcal{U}$ measures the temporal average of the maximal local joint velocity. We then use the factor $\mathcal{U}^s/\mathcal{U}^c$ to warp the global velocity of the root joint, where the $\mathcal{U}^s$ and $\mathcal{U}^c$ represent the velocity factors for the style example $m^s$ and content motion $m^c$, respectively. 

\section{Implementation Details}\label{ssec:implementation}
We use the Pytorch platform and a single 3090Ti graph card for the pre-training and fine-tuning with a batch size of 64. \orange{We employ the AdamW~\cite{loshchilov2018decoupled} optimizer with $\beta_1=0.9, \beta_2=0.999$. The learning rate is set to 0.0001 for all models. When fine-tuning, the} $\lambda_{sr}$ and $\lambda_{s}$ are set to 1 and 0.1, respectively. The generative prior $\varepsilon_{\theta}$ and fine-tuned style transfer model $\varepsilon_{\theta s}$ have the same Transformer encoder architecture which consists of 8 Transformer layers with 512 latent sizes and 1024 feedforward dimensions. Motion inputs to $\varepsilon_{\theta}$, $\varepsilon_{\theta s}$, or the motion-semantic discriminator $Dis$ need to first pass through a shared embedding layer to encode the raw motion feature $m$ into 512-dimensional deep features. In contrast, the output layer in $\varepsilon_{\theta}$ and  $\varepsilon_{\theta s}$ is the inverse of input embedding, which decodes the 512-dimensional features into the raw motion representation. The layer number $N$ of the motion-semantic discriminator $Dis$ is set to 8. When generating the style-neutral motion pairs in Sec~\ref{ssec:fine-tune}, \orange{we experimentally find the noisy step $G=950, K=300$ to be an appropriate choice (discuss in  Sec~\ref{ssec:K_and_G}).}

\orange{At each optimization step in fine-tuning, we need to denoise in a loop to get the stylized motion (see Equation~\ref{eq:style_denoising}).} To accelerate the procedure and make the style transfer stable, we use deterministic DDIM like~\cite{kim2022diffusionclip} during the forward and reverse processes of  $\varepsilon_{\theta s}$. Specifically, We perform the total sample steps S=20 in DDIM compared to the T=1000 steps in DDPM. Thus, K = 300 corresponds to 6 steps in the actual accelerated process. However, we still use DDPM when generating the style-neutral motion pair. \orange{We fine-tune the model using neutral motion data from the text-motion training dataset as well as style-neutral motion pair <$m^s, m^{s\_n}$> for one epoch.} For more information about pre-training and fine-tuning time efficiency, please refer to the Appendix.

\section{Experiments and Evaluation}
\begin{figure*}
    \centering \includegraphics[width=\linewidth]{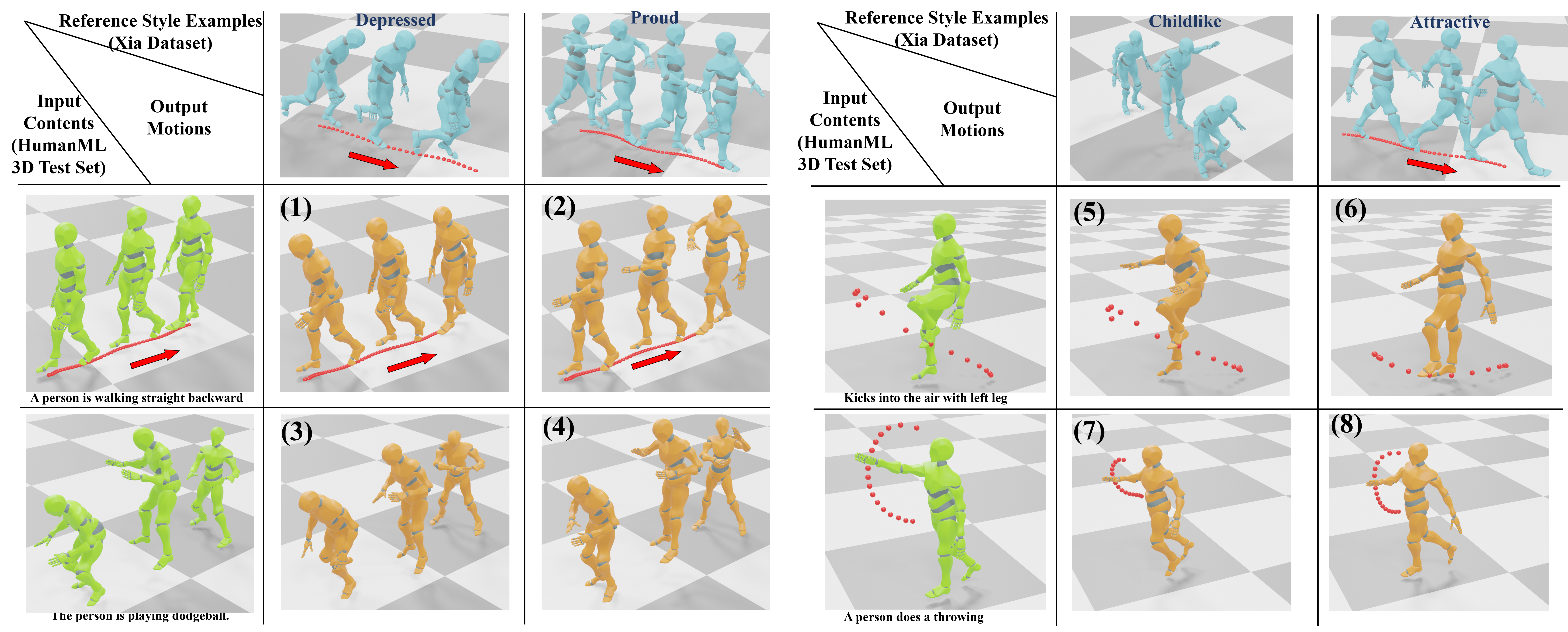}
    \caption{\orange{Qualitative results of our few-shot style transfer. We can transfer unseen motion styles} (from Xia dataset) to different content motions (from HumanML3D test set).}
    \label{fig:unseen_examples}
    
\end{figure*}

\begin{figure}[t]
    \centering
    \includegraphics[width=\linewidth]{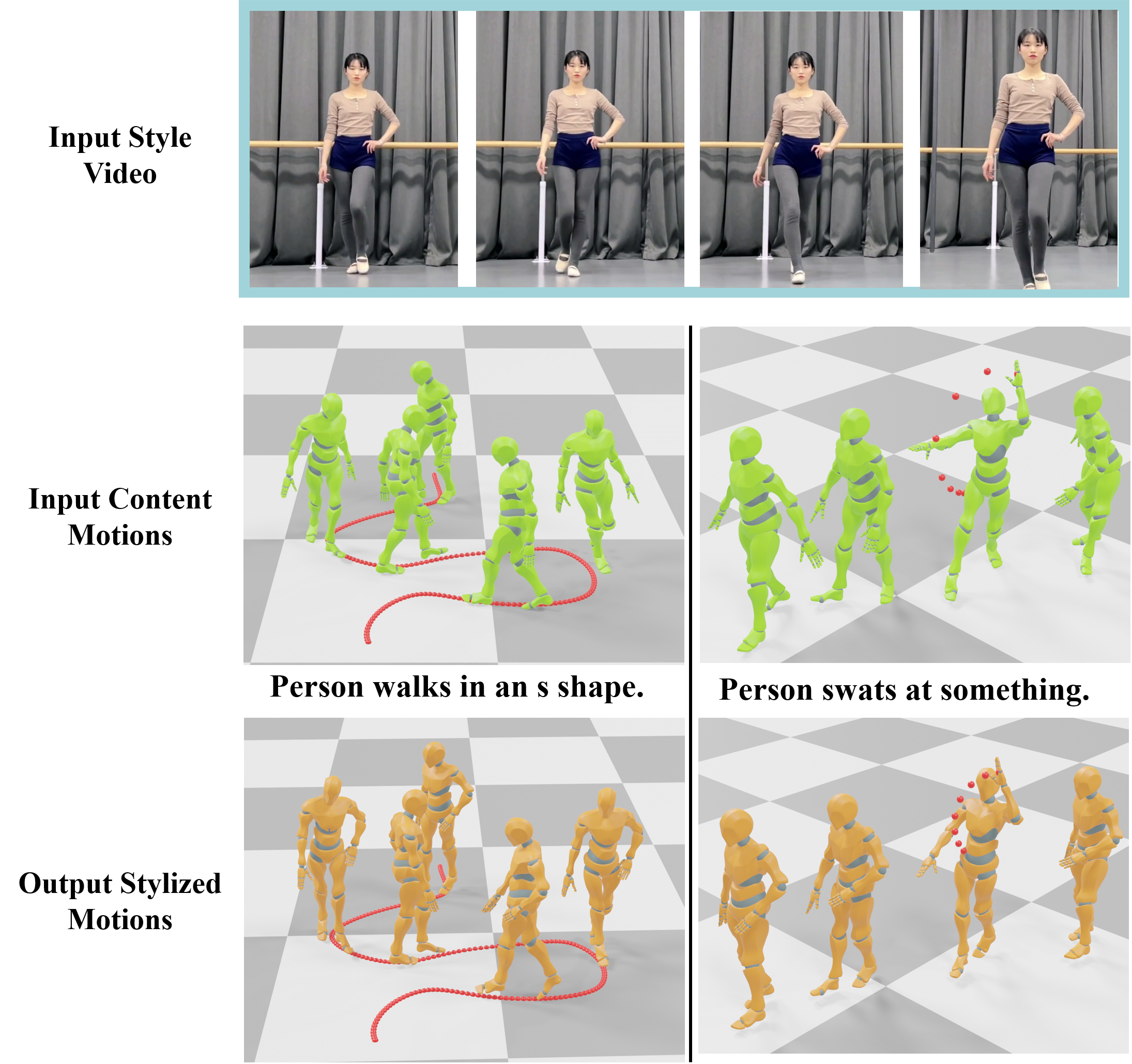}
    \caption{\orange{Qualitative results. We can learn} the human motion style from video and transfer it to different content motions.}
    \label{fig:video_style}
\end{figure}

In Sec~\ref{ssec:unseen_experiments}, we first qualitatively assess the ability of our method to transfer unseen style to diverse motion contents given only a segment of style motion example. Next, in Sec~\ref{ssec:comparisons_sota}, we quantitatively and qualitatively compare our method with state-of-the-art approaches on the dedicated style motion datasets (with content and style annotation).

\subsection{Transfer Unseen Style to Diverse Content Motions}\label{ssec:unseen_experiments}
In this part, we design experiments to demonstrate that our two-stage framework can acquire the generative ability from large-scale text-motion datasets and then be transformed into a style transfer model ($\varepsilon_\theta \rightarrow \varepsilon_{\theta s}$) by fine-tuning with a single style example. The datasets we used are as follows:

\textbf{HumanML3D}~\cite{Guo_2022_CVPR} is a stylistically unbalanced (predominantly neutral), text-annotated large 3D human motion dataset. It encompasses a wide range of human contents extracted from the AMASS~\cite{AMASS:ICCV:2019}, including daily activities (e.g. walking and jumping), sports, and artistic pursuits such as dancing. We follow the official train-test split.  

\textbf{Xia dataset}~\cite{xia2015realtime} is a dedicated style motion dataset (with content/style labels), which has been widely used in previous motion style transfer work~\cite{aberman2020unpaired, park2021diverse, jang2022motion, song2023finestyle}. It comprises a total of 79,829 frames of motion data and consists of 6 motion contents (walk, run, jump, etc.) and 8 styles (angry, childlike, depressed, etc.). We down-sample the framerate of the Xia dataset to be consistent with HumanML3D and retarget the motion to the SMPL~\cite{loper2023smpl} skeleton structure. Since this dataset does not contain textual annotations, we simply annotate the motion with text based on the content and style labels. For example, an annotation might read 'a person is running in depression'.

In stage \uppercase\expandafter{\romannumeral1}, we pre-train the text-to-motion generative prior $\varepsilon_\theta$ on the HumanML3D train set. In stage \uppercase\expandafter{\romannumeral2}, we sample an unseen style example $m^s$ from the Xia dataset for style fine-tuning ($\varepsilon_\theta \rightarrow \varepsilon_{\theta s}$). At inference, we input content motions from the HumanML3D test set to be converted to the style corresponding to $m^s$. 

It is difficult to quantitatively evaluate the style transfer performance due to the domain differences between the two datasets. Therefore, we qualitatively evaluate our method as shown in Figure~\ref{fig:unseen_examples}. The results (1)(2) show that different motion styles (depressed, proud) can be transferred to the same backward motion, despite the style example containing only forward movements. The transferred motions (3)(4) demonstrate that our method can well preserve the action of the content motions while expressing the appropriate stylistic characteristics of the style examples. The results (5)$\sim$(8) demonstrate the ability of our method to handle the style transfer between heterogeneous motions, where the given style examples are locomotion and the content motions are non-locomotion. Notably, the stylized output motions exhibit distinct end-effector trajectories compared to the content motions, with the trajectory morphology and body pose collectively conveying the corresponding style.

\textbf{Transfer style from video.} Given our framework's capability to fine-tune using just one style example, we are prompted to leverage a diverse array of video resources containing stylized human movements as style exemplars. Specifically, we first use HybrIK~\cite{li2021hybrik} to extract the human poses in the video, and then apply a Butterworth low-pass filter to remove the jitter from the estimated motion. The cut-off frequency of the filter is set to 3, and we exclude the 3D global translation of the estimated motion during this process. Then, we proceed with the fine-tuning stage pipeline to derive a transfer model $\varepsilon_{\theta s}$ associated with the video. Subsequently, the style transfer diffusion model $\varepsilon_{\theta s}$ can apply the human style of the video to diverse content motions. As depicted in Figure~\ref{fig:video_style}, we observe that the walking style from the video can be applied to other walking trajectories, such as the S-shape. Furthermore, although the akimbo is considered part of the motion style in the video, the semantics of the content motion are effectively preserved when waving the left hand. For more qualitative results, please refer to our supplementary video. 

\begin{figure}
    \centering
    \includegraphics[width=\linewidth]{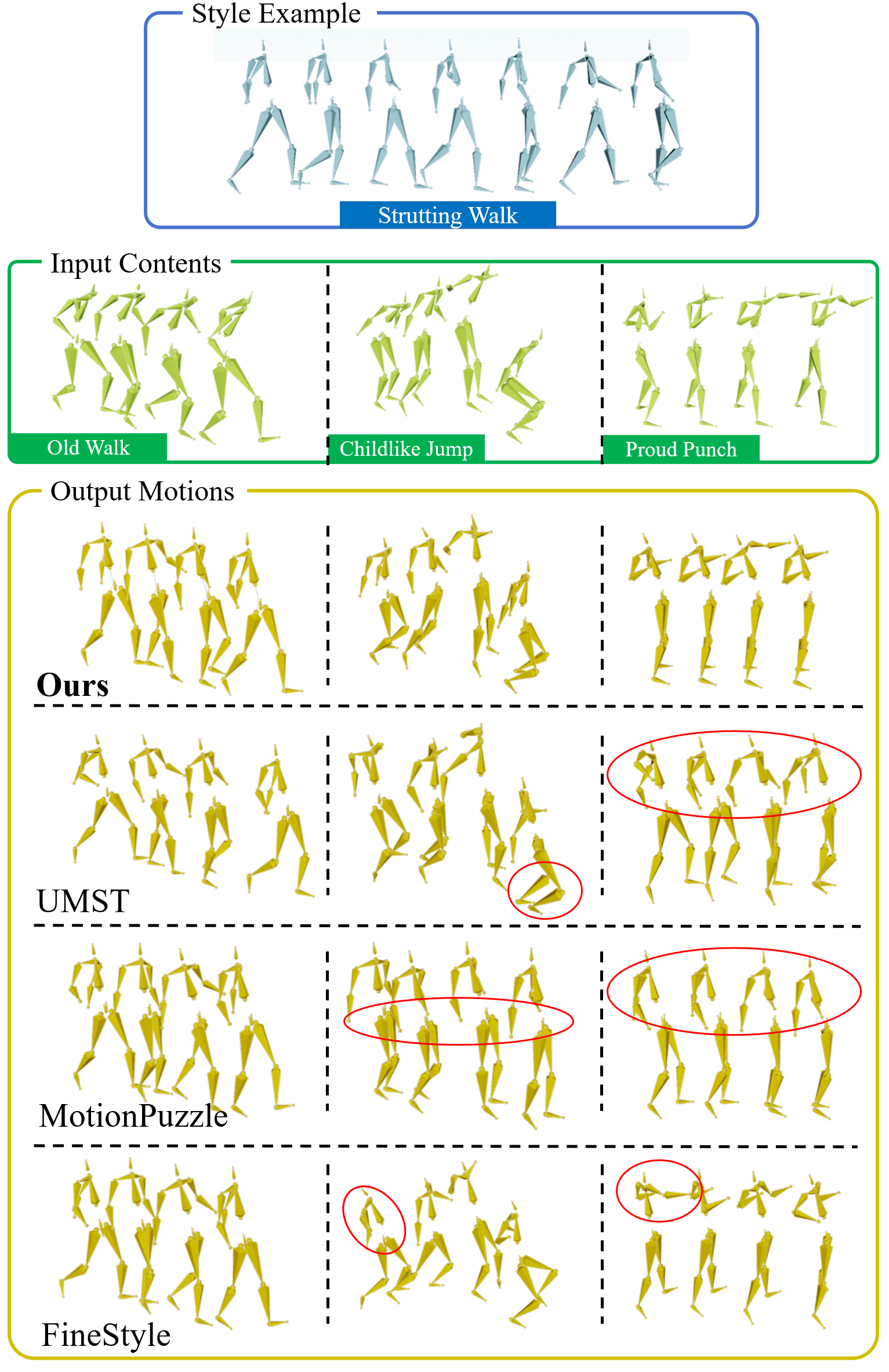}
    \caption{Qualitative \orange{comparison with state-of-the-art methods on Xia test set.}}
    \label{fig:qualitative_comparison}
\end{figure}

\subsection{Comparisons with SOTA}\label{ssec:comparisons_sota}
In this section, we quantitatively and qualitatively compare our style transfer method with UMST\cite{aberman2020unpaired}, MotionPuzzle\cite{jang2022motion}, and FineStyle\cite{song2023finestyle}. \orange{The implementation of these methods follows their published code.}
\begin{table*}[h]
    \centering
    \caption{Quantitative comparison with state-of-the-art methods. \orange{FMD is short for  Fréchet Motion Distance. CRA and SRA stand for Content Recognition Accuracy and Style Recognition Accuracy, respectively. The down arrow indicates that smaller values are better and vice versa.}}
    \begin{tabular}{c|c|c|c||c|c|c}
        \hline
         \multirow{2}{*}{Methods}& \multicolumn{3}{c||}{\orange{Xia~\cite{xia2015realtime}} } & \multicolumn{3}{c}{\orange{Bandai-2\cite{kobayashi2023motion}}}\\
         \cline{2-7}
          & FMD$\downarrow$ & CRA$\uparrow (\%)$ & SRA$\uparrow$ (\%)& FMD $\downarrow$& CRA$\uparrow$(\%) & SRA$\uparrow$(\%) \\
         \hline
         UMST~\cite{aberman2020unpaired} &5.63 & 62.88 & 38.97 & 23.38 & 53.78 &54.23 \\
         MotionPuzzle~\cite{jang2022motion} & 22.79 & 40.19 & 22.56 & 38.63 & 42.67 & 25.79 \\ 
         FineStyle~\cite{song2023finestyle}  & 5.34 & 63.91 & 55.58 & 24.84 & 58.89 & 56.01 \\ 
         Ours & \textbf{3.84} & \textbf{83.72} & \textbf{70.77} & \textbf{2.33} & \textbf{93.41} & \textbf{60.11}\\
         \hline
    \end{tabular}
    
    \label{tab:comparison_sota}
    
\end{table*}

\textbf{Quantitative evaluation.} To quantitatively compare our style transfer framework with state-of-the-art methods. we conducted experiments on dedicated style motion datasets: Xia and Bandai-2~\cite{kobayashi2023motion}, respectively. The Bandai-2 dataset is similar in composition to Xia but has a larger scale, containing 10 content and 7 styles totaling 384,931 frames of motion. For the Xia dataset, we follow~\cite{song2023finestyle} to split the train-test set. For the Bandai-2, we uniformly select test samples from each content and style, resulting in a total of 3780 transfer combinations.  

Since our approach is two-stage, for a fair comparison, \orange{we pre-train and fine-tune the models ($\varepsilon_\theta$, $\varepsilon_\theta \rightarrow \varepsilon_{\theta s}$ ) under the same data conditions (Xia or Bandai-2 training set) as the baselines. Meanwhile, the baselines will also fine-tune the networks with the input style example for the same number of steps before testing.} We follow~\cite{jang2022motion} and use three metrics: Content Recognition Accuracy (CRA), Style Recognition Accuracy (SRA), and Fréchet Motion Distance (FMD) to assess the content preservation, style accuracy, and quality of the transferred human motions, respectively.

To compute the CRA, we pre-train a content classifier~\cite{yan2018spatial} to categorize the generated motions and examine whether they match the label of the source content motions. Similarly, we pre-train another style classifier to categorize the style of the generated
motions. The FMD, a derivative of the Fréchet Inception Distance (FID)~\cite{heusel2017gans}, can evaluate the resemblance between generated and real motions by comparing their deep feature distributions. In accordance with~\cite{jang2022motion}, we employ the \orange{above} content classifier to extract the latent features from the classifier's final pooling layer and calculate the FMD score. 

The quantitative results are presented in Table~\ref{tab:comparison_sota}. Our method achieves state-of-the-art performance on both the Xia and Bandai-2 test sets. From the table, our CRA significantly outperforms other methods, achieving 83.72\% and 93.41\% accuracy on the Xia and Bandai-2 datasets, respectively.  Regarding motion quality, our method not only yields the lowest FMD scores but also remains unaffected by the dataset scale (3.84 and 2.33 on the Xia and Bandai-2 datasets, respectively). Furthermore, our approach achieves the highest SRA on both datasets, demonstrating a robust balance between content preservation and style transfer. \orange{The high performance of our method may be attributed to the explicit definition of style as a particular variation of motion and adding it through the denoising process, which is more interpretive and stable. Also, the style transfer guided by text prompts has more detailed information compared to style labels.}

\textbf{Qualitative evaluation.} 
To evaluate the performance of style transfer on heterogeneous motions, we showcase the qualitative results on the Xia test set in Figure~\ref{fig:qualitative_comparison}. The input style example is a clip of \textit{"strutting walk"}, while the content motions encompass \textit{"old walk", "childlike jump" and "proud punch"}. In the first column, where the content motion shares the same action type (i.e., walking) as the style example, all methods successfully execute style transfer from \textit{"old walk"} to \textit{"strutting walk"}. In the second column, MotionPuzzle fails to persevere the source content, yielding a drifting motion, while both UMST and FineStyle generate motions with unnatural poses. These samples show that our approach achieves robust style transfer and content preservation even among challenging heterogeneous motions. For more qualitative results, please refer to the Appendix and supplementary video.

\subsection{Visualization by t-SNE}

\begin{figure}
    \centering
    \includegraphics[width=\linewidth]{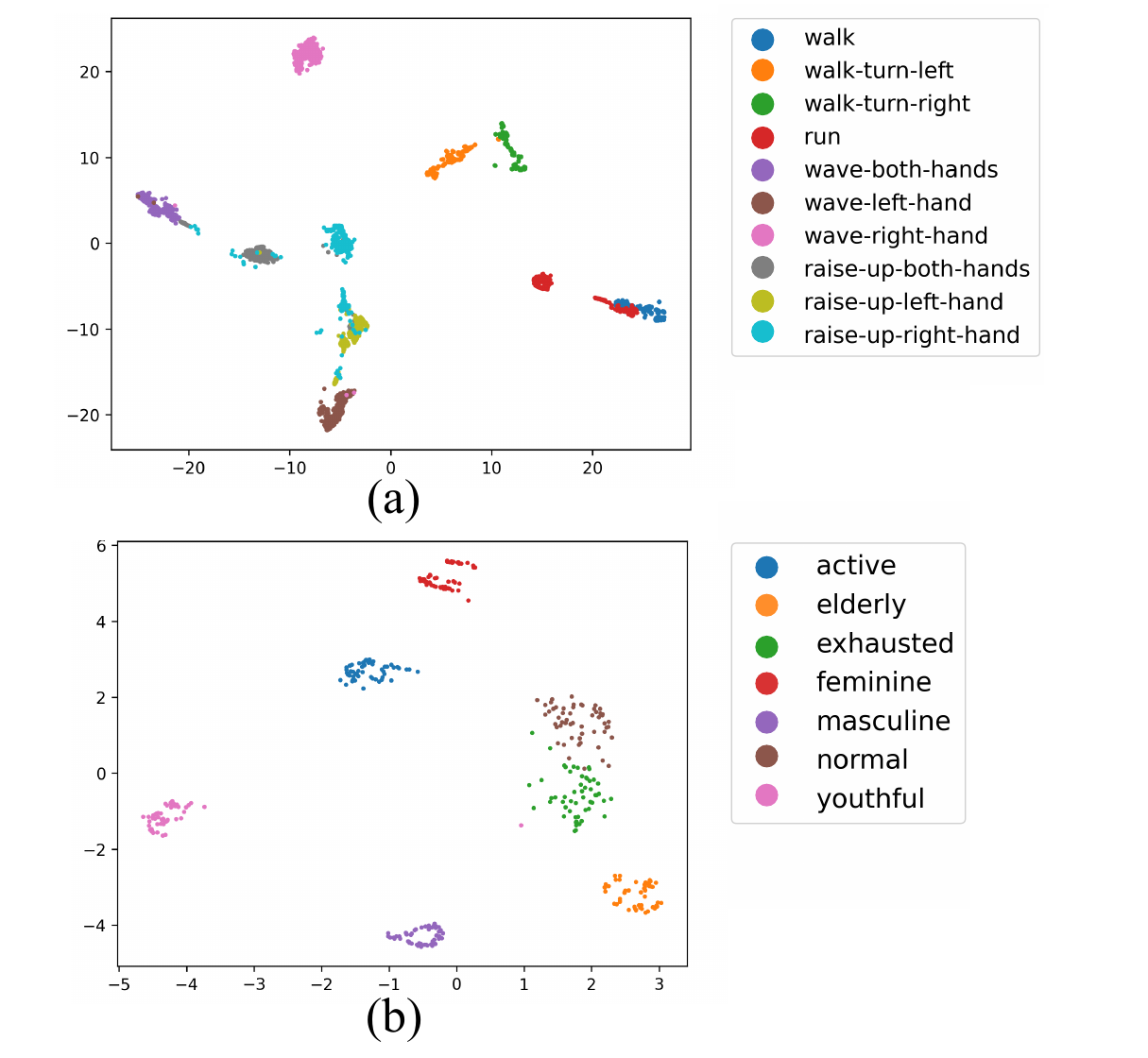}
    \caption{Hidden feature visualization via t-SNE for our transferred motions based on (a) content labels and (b) style labels.}
    \label{fig:tsne}
\end{figure}
To intuitively visualize the comprehensive outcomes of our style transfer, we employ t-distributed stochastic neighbor embedding (t-SNE)~\cite{van2008visualizing} to illustrate the clustering efficacy of the transferred motions. Specifically, we project the hidden features, extracted by the final pooling layer of content and style classifiers~\cite{yan2018spatial}, into a two-dimensional space. This experimental analysis is conducted on the Bandai-2 test set. As depicted in Figure~\ref{fig:tsne}(a), the hidden features of our stylized motions exhibit discriminative cluster boundaries based on the content labels. Furthermore, these clusters demonstrate the proximity relationships between motion contents, with similar motions grouped closely together. 

Recognizing that identical styles may manifest differently across heterogeneous motion contents, we project the stylized motions with the \textit{"raise-up-both-hands"} content into a two-dimensional space and color them according to the style labels. As shown in Figure~\ref{fig:tsne}(b), the clustering reveals that the output stylized motions exhibit clear style distinctions and can be effectively differentiated by the classifier. In summary, the clustering results indicate that our stylized motions possess adequate discriminative features, thereby further validating the effectiveness of our approach.

\section{Ablation Study}\label{ssec:ablation}
We conducted the ablation study, with a primary focus on three key aspects: the selection of noise steps, the reliability of neutral motion generation, and the key components of the framework.
\begin{figure}
    \centering
    \includegraphics[width=\linewidth]{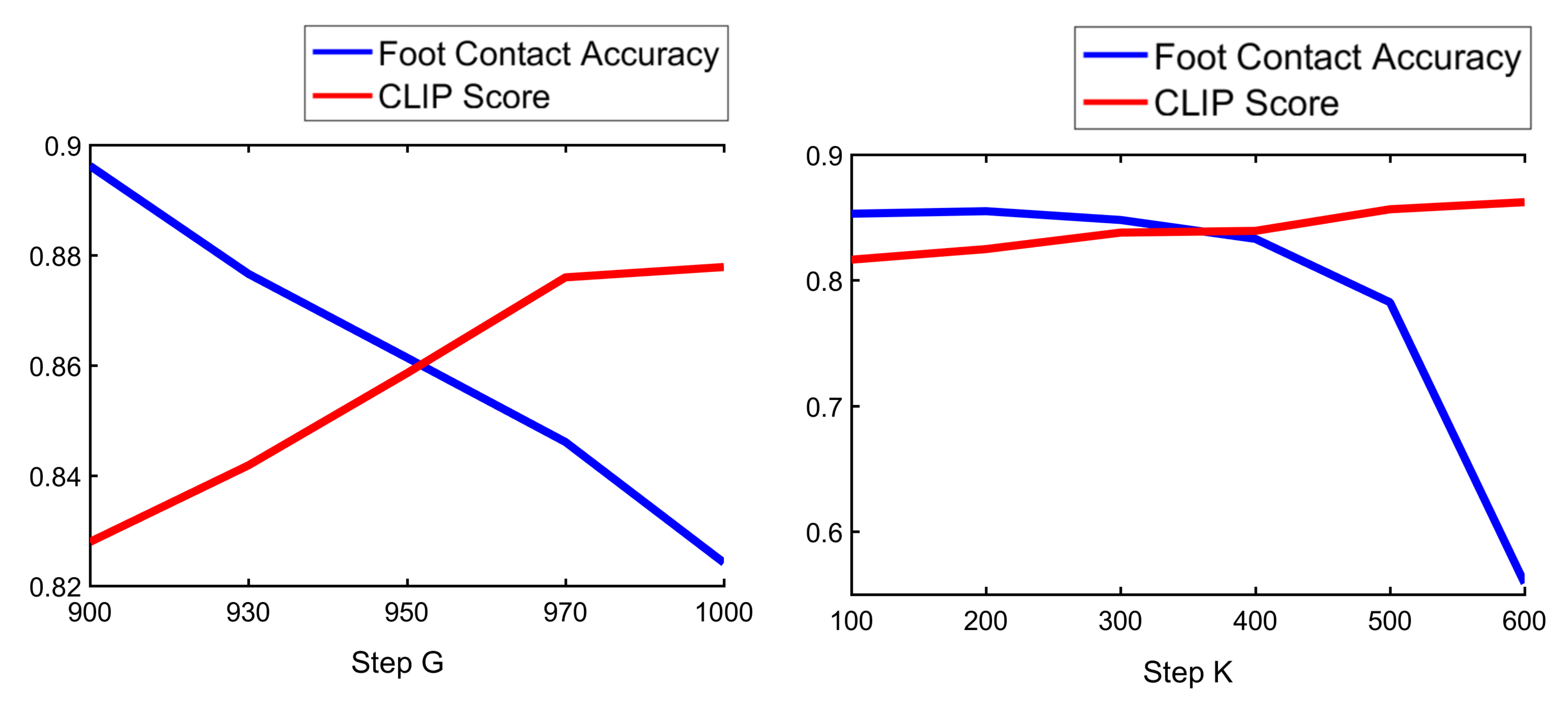}
    \caption{Ablation study 
 \orange{on noise step $G$ and $K$}}
    \label{fig:GK}
\end{figure}

\subsection{Selection of Noise Steps $K$ and $G$}\label{ssec:K_and_G}
We seek to find the most appropriate noise step $G$ during neutral motion generation and the noise step $K$ during style transfer.

\orange{According to Equation~\ref{eq:noise_step}}, a smaller \orange{noise step $G$} during the forward process preserves more detail in the style example $m^s$, while a larger $G$ will \orange{make the noised motion $m^s_G$ closer to the standard normal distribution, thus encouraging the motion generated by Equation~\ref{eq:mdm_denoising} to be guided by a neutral textual prompt. Therefore, we wish to find a proper $G$ such that the generated motion retains the content of $m^s$ while being stylistically neutral.} We use the Foot Contact Accuracy~\cite{villegas2021contact} to measure the degree of alignment between the $<m^s, m^{s\_n}>$\orange{, where $m^{s\_n}$ is the neutral motion corresponding to $m^s$ (defined in Sec~\ref{ssec:fine-tune}).} On the other hand, we measure neutrality by comparing the similarity of the generated neutral motions $m^{s\_n}$ with the neutral textual descriptions via the motion-semantic discriminator $Dis$ in the CLIP space. Similarly, we select $K$ with the same metrics to find the balance between content preservation and style transfer.

The evaluations are conducted on the Bandai-2 test set and the results are presented in Figure~\ref{fig:GK}. In the left subplot, we can see that the metrics reach a balance around G=950. Similarly, from the right subplot, it is observed that as the number of $K$ increases, the CLIP score increases accordingly, indicating that the style of the generated motion is getting closer to the input style examples. However, after 400 steps, the Foot Contact Accuracy of the generated stylized motion starts to degrade significantly. Therefore, we chose $K = 300$ as the number of noising/denoising steps in the forward/reverse process during the style transfer diffusion. 

\begin{table}[t]
    \centering
    \caption{Ablation study of style-neutral motion pair generation on Bandai-2 test set. }
    \begin{tabular}{c|c|c|c}
        \hline
         \multirow{2}{*}{Methods} & \multicolumn{3}{c}{Ablation on Paired Neutral motion Generation}\\
         \cline{2-4}   
         & FMD$\downarrow$ & CRA$\uparrow (\%)$  & SRA$\uparrow$ (\%)\\
         \hline
         Ours & \textbf{5.87} & 98.06 & 88.61 \\
        \hline
        KNN+DTW & 21.76 & \textbf{100} & \textbf{97.78}  \\
        \hline
    \end{tabular}
    
    \label{tab:ablation_neutral_generation}
    
\end{table}

\begin{table}[t]
    \centering
    \caption{Ablation study of the key components of the framework on Xia test set.}
    \begin{tabular}{c|c|c|c}
        \hline
         \multirow{2}{*}{Methods} & \multicolumn{3}{c}{Ablation on Style Transfer}\\
         \cline{2-4}   
         & FMD$\downarrow$ & CRA$\uparrow (\%)$  & SRA$\uparrow$ (\%)\\
         \hline
        Ours & \textbf{3.84} & 83.72 & \textbf{70.77} \\
        \hline
        with KNN+DTW & 5.48 & 76.86 & 69.49 \\
 
        with AE & 5.84 & 19.68 & 13.27 \\
        without $\mathcal{L}_s$ & 4.00 & \textbf{88.01} &28.01  \\
        \hline
    \end{tabular}
    \label{tab:ablation_structure}
    
\end{table}
\subsection{The Reliability of Style-Neutral Motion Pairs}
To demonstrate that the paired neutral motions generated through our prior model $\varepsilon_\theta$ are reliable, we compare our generated results with the neutral motions produced by KNN+DTW. Specifically, given an style example $m^s$, we generate $m^{s\_n}$ using the method described in Sec~\ref{ssec:fine-tune}, whereas KNN+DTW first searches for a segment of neutral motion (has the same content label as $m^s$) from the dataset that is closest to $m^s$, and then aligns it to $m^s$ via DTW. The comparison results are illustrated in Table~\ref{tab:ablation_neutral_generation}. Since KNN+DTW searches directly from the dataset, the obtained motions can be easily recognized by the classifier, leading to high CRA as well as SRA. However, the use of DTW alignment compresses or stretches the neutral motions along the temporal axis, leading to an increase in FMD. Our method, on the other hand, is able to achieve a relatively balanced and competitive performance. 

In addition to directly evaluating the generated neutral pairs, we also test the effect on the final style transfer, the results are shown in Table~\ref{tab:ablation_structure}. our method is able to achieve better style transfer performance with a marginal margin compared to KNN+DTW. For more qualitative results, please refer to our Appendix and supplementary video.

\subsection{Key Components of the Framework}

\noindent\textbf{Effect of diffusion model.}
To validate the effectiveness of the diffusion model, we replace it with an autoencoder(AE) architecture in both the pre-training and fine-tuning stages. Specifically, we adopt the AE architecture in UMST, while omitting the Instance Normalization and AdaIN layers. From Table~\ref{tab:ablation_structure}, we observe the performance of our model with AE architecture decreases in all three metrics. This issue may arise from the AE model being trained to over-fit the motion reconstruction during stage \uppercase\expandafter{\romannumeral1}, making it difficult to achieve effective style transfer through a few-shot fine-tuning. 

\noindent\textbf{Effect of $\mathcal{L}_s$.} The semantic-guided style transfer learning loss $\mathcal{L}_s$ enables us to supervise the fine-tuning in CLIP semantic space. To validate the effectiveness, we remove the loss $\mathcal{L}_s$ during the fine-tuning stage. It is observed the CRA in Table~\ref{tab:ablation_structure} of the model increases from 83.72\% to 88.01\% and the SRA decreases dramatically from 70.77\% to 28.01\%. This is because, without $\mathcal{L}_s$, the model's ability to transfer style can only be learned by examining differences in style-neutral motion pair $<m^s, m^{s\_n}>$, and thus is difficult to generalize to heterogeneous motions. Indeed, the $\mathcal{L}_s$ fully exploits the motion style prior in the text-to-motion dataset, which allows the model $\varepsilon_{\theta s}$ to learn style transfer both from the example $m^s$ and from the motions with the same style type as $m^s$. 
\section{Discussion}
In this paper, we present a novel two-stage framework for human motion style transfer. In stage \uppercase\expandafter{\romannumeral1}, we pre-train a diffusion-based T2M model $\varepsilon_\theta$ as generative prior, which allows us to adapt to heterogeneous motions. We realize that noise in the reverse process can bring about variations in the output motion, and human motion styles, as a particular variation, can also be generated in the reverse process. Therefore, in stage \uppercase\expandafter{\romannumeral2}, we fine-tune the diffusion model based on the input style example, making the reverse process of $\varepsilon_{\theta s}$ has the capability of style transfer. This two-stage framework allows us to leverage large, stylistically unbalanced text-motion datasets to train a robust generator, while in the fine-tuning for style transfer, we only need a segment of the style example. Extensive experiments demonstrate that our method achieves state-of-the-art performance and has practical applications. 

Our system has two main limitations. First, our style transfer model takes the whole motion segment as the input rather than the sliding window, which makes it difficult to combine with real-time motion control methods like PFNN~\cite{holden2017phase}. In the future, improving the denoising efficiency and combining the motion phase~\cite{starke2022deepphase} may be a solution to achieve diffusion-based real-time style transfer. 

Second, we fine-tune an independent style transfer model for each different style example, which may result in model storage stacking \orange{as well as difficulties in style interpolation}. In future work, we can introduce new conditions (e.g. the hidden features of the style example) in the reverse process of style diffusion to integrate knowledge from different style transfer models. \orange{We can also consider introducing the DiffusionBlending technique in PriorMDM~\cite{shafir2024human} for style interpolation and extrapolation.}




\subsubsection*{Acknowledgment}
This work was supported by National Key R\&D Program of China (NO. 2022YFB3303202).

\bibliographystyle{eg-alpha-doi} 
\bibliography{main}  


\newpage


\end{document}